# An Intersectional Analysis of Long COVID Prevalence

**Authors**
Jennifer Cohen[1,2] and Yana van der Meulen Rodgers[3]

    1. Department of Global and Intercultural Studies, Miami University, Oxford, OH, USA

    2. Ezintsha, Wits Reproductive Health and HIV Institute, Department of Medicine, Faculty of Health Sciences, University of the Witwatersrand, Johannesburg, South Africa

    3. Department of Labor Studies and Employment Relations, Rutgers University, New Brunswick, NJ, USA

**Abstract**
    Background

        Long COVID symptoms – which include brain fog, depression, and fatigue – are mild at best and debilitating at worst. Some U.S. health surveys have found that women, lower income individuals, and those with less education are overrepresented among adults with long COVID, but these studies do not address intersectionality. To fill this gap, we conduct an intersectional analysis of the prevalence and outcomes of long COVID in the U.S. We posit that disparities in long COVID have less to do with the virus itself and more to do with social determinants of health, especially those associated with occupational segregation and the gendered division of household work.

    Methods

        We use 10 rounds of Household Pulse Survey (HPS) data collected between June 2022 and March 2023 to perform an intersectional analysis using a battery of descriptive statistics that evaluate (1) the prevalence of long COVID and (2) the interference of long COVID symptoms with day-to-day activities. We also use the HPS data to estimate a set of multivariate logistic regressions that relate the odds of having long COVID and activity limitations due to long COVID to a set of

individual characteristics as well as intersections by sex, race/ethnicity, education, and sexual orientation and gender identity.

Results

Findings indicate that women, some people of color, sexual and gender minorities, and people without college degrees are more likely to have long COVID and to have activity limitations from long COVID. Women have considerably higher odds of developing long COVID compared to men, a disparity exacerbated by having less education. Intersectional analysis by gender, race, ethnicity, and education reveals a striking step-like pattern: college-educated men have the lowest prevalence of long COVID while women without college educations have the highest prevalence. Daily activity limitations are more evenly distributed across demographics, but a different step-like pattern is present: fewer women with degrees have activity limitations while limitations are more widespread among men without degrees. Regression results confirm the negative association of long COVID with being a woman, less educated, Hispanic, and a sexual and gender minority, while results for the intersectional effects are more nuanced.

Conclusions

Results point to systematic disparities in health, highlighting the urgent need for policies that increase access to quality healthcare, strengthen the social safety net, and reduce economic precarity.

## I. Introduction

Emerging evidence suggests that social determinants of health play a role in COVID-19 exposure and infection rates, but less is known about the prevalence of post-acute sequelae of SARS-CoV-2 (henceforth long COVID) and associated activity limitations. [1] **The U.S. Centers for Disease Control has defined long COVID formally as "a range of new, returning, or ongoing health problems lasting four or more weeks after first being infected with COVID-19." [2]** Long COVID symptoms – which include fatigue, post-exertional malaise, memory loss, and other neurocognitive impairments – are mild at best and debilitating at worst. Some U.S. health surveys have found that women, lower income individuals, and those with less education are overrepresented among adults with long COVID. [3-6] However, these studies do not adequately address intersectionality, nor do they have a conceptual framework to understand why long COVID would disproportionately impact sub-populations that have historically been economically and socially marginalized.

Intersecting social determinants of health – especially those related gender, race, class, and sexuality – are increasingly recognized as multifactorial contributors to morbidity and mortality in medical research. Low socioeconomic status, sexism, racism, and heteronormativity imply varying degrees and forms of precarity, discrimination, and distress. [7, 8] Intersectionality originated as a critique of analytical approaches that treat multiple, interacting forms of oppression as mutually exclusive or additive rather than interactive, thereby creating social location-specific experiences of marginalization for the multiply-subordinated. [9] Using a lens of intersectionality means viewing race, gender and class as interwoven systems of oppression rather than the sum of individual experiences. [10, 11] Intersectionality locates complex group-level inequalities across multiple, intersecting hierarchical social relationships with the theoretical aim of clarifying the structural determinants of those inequalities. Intersectionality is notoriously difficult to capture in statistical analyses because of its complexity, but it is evident in health outcomes. [12-17] Like other outcomes, long COVID is likely to be unevenly distributed across populations by gender, income, education, race/ethnicity, and sexual and gender minority (SGM) status.

Racism and sexism are critically important determinants of health. [18] Notably, it is *racism,* not race, and *sexism*, not gender, that are social and structural determinants of health. Social determinants of health are the systems of social relations, or social processes, that reinforce hierarchies of constructed categories to the detriment of people at the bottom of those hierarchies – women, people of color, sexual and gender minorities, and those with disabilities – and to the benefit of people at the top. An intersectional approach requires consideration of multiple, interacting hierarchies and their origins. [19] The categorical approach to intersectionality is necessarily comparative and multigroup. [12,13] In categorical intersectional analyses, demographic variables serve as proxies for hierarchical power relations. Variation across groups, including disparities in health, largely reflect social dynamics and relations of power. Social determinants of health are widely acknowledged as primary determinants of health that encompass inequities resulting from social structures. [18]

Several theoretical frameworks allow intersectional analyses in health. In particular, the psychosocial approach to social determinants of health emphasizes biological responses to social interactions, especially to stress, and is used in studies of allostatic load. [20] In fundamental cause theory, social inequalities observed downstream are the consequence of fundamental inequalities upstream. [21, 22] The political economy of health has materialist foundations; political and economic forces that generate and reproduce inequality are root causes of social inequalities in health. [20] This approach clarifies effects of institutional and structural forms of power, economic and otherwise.  Embracing the political economy approach is Krieger's comprehensive multiscalar ecosocial theory, which focuses on population-level dynamics, but retains biology as a feature while discarding assumptions of biomedical individualism. [19, 20] This framework engages with biological manifestations of social relations, and embodiment – how humans integrate the material and social world into their biology – plays a key conceptual role. [20]  Pathways to embodiment are simultaneously structured by trajectories of biological and social development, the elements of which clarify the multi-level pathways that connect biological embodiment with sexism, racism, and other hierarchical social relations. The ecosocial approach integrates discrimination and material forces that result in inequality and inequity, and thus guides our study.

We hypothesize that environmental factors that contribute to other forms of morbidity among those who are socially and economically marginalized also make these sub-populations more vulnerable to persistent COVID symptoms and novel outcomes related to COVID that can continue for months. To test this hypothesis, we perform an intersectional analysis using a battery of descriptive statistics as well as multivariate logistic regressions applied to data from the U.S. Census Bureau's Household Pulse Survey (HPS).

## II. Data and Methods

### A. Data

Our analysis evaluates (1) the prevalence of long COVID and (2) the impact of long COVID on day-to-day activities across demographic groups. The HPS is a rapid deployment, rapid dissemination survey about COVID and other emergent issues to inform federal and state government responses. It is a biweekly survey of between 40,000 and 75,000 individuals. The Census Bureau added questions about long COVID to the HPS in June 2022. We use the 10 rounds of HPS data collected between June 2022 to March 2023 to track prevalence of long COVID and extent of related activity limitations.

The HPS question about COVID infection reads, "Have you ever tested (using a rapid point-of-care test, self-test, or laboratory test) positive for COVID-19 or been told by a doctor or other health care provider that you have or had COVID-19?" Survey questions about long COVID ask "Did you have any symptoms lasting 3 months or longer that you did not have prior to having coronavirus or COVID-19?" and "Do these long-term symptoms reduce your ability to carry out day-to-day activities compared with the time before you had COVID-19?" The 3-month duration of persistent symptoms as indicative of long COVID symptoms is used in several other studies, including the REACT study in England. [23] Note that the survey does not ask if people had long COVID; rather, it asks about persistent symptom and identifies a range of examples of symptoms (including fatigue, difficulty concentrating, forgetfulness, shortness of breath, joint pain, and dizziness). So, respondents need not know anything about long COVID to answer the question to the best of their ability. The question on day-to-day activity limitation did not offer any details about specific activities, such as work, socializing, and mobility, that might be impacted by symptoms.

The survey also collects data on gender identity, sex assigned at birth, race/ethnicity, sexuality, level of education, and age. Data on race is collected separately from ethnicity, hence the baseline data are presented as white/non-white and as Hispanic/non-Hispanic. Like non-Hispanic respondents, Hispanic survey respondents may identify as white, Black, Asian, or multiracial/other. The transgender category groups trans women and trans men together. "None of these" – meaning "female, male, transgender" – is a category that presumably includes non-binary, agender, genderfluid, and other gender identities. We refer to this group as genderqueer. The imposition of an SGM category allows us to examine differences between people who are SGM and cisgender, straight people. This combined category is heterogeneous; we do not mean to imply that gay, bisexual, transgender, and genderqueer people have comparable experiences in daily life or in the labor market. Education is our key indicator of an individual's socioeconomic status.

### B. Methods

In the descriptive analysis we report average prevalence for 2022 (7 survey rounds between June 2022 and December 2022), and we report impact on activities for all periods in which the question was asked (7 rounds between September 2022 and March 2023). The sample consists of novel cohorts and is pooled. In our descriptive analysis of prevalence, sample means are conditional on ever having had a COVID infection prior to the survey. By mid-November 2022, 50% of American adults reported being infected, with 35% saying they had tested positive. [24] The descriptive analysis uses the gender identity question ("female, male, transgender, none of these") for the gender analyses. Our focus on the development of long COVID among those who have ever been infected means that our results reflect an individual's vulnerability to longer-term adverse health outcomes from having COVID, not the likelihood of an initial exposure to COVID. That said, we conducted robustness checks using the full sample of adults, not only those who have had COVID, and our substantive conclusions do not change. In our descriptive analysis of activity limitations, sample means are conditional on ever having had long COVID. In this way, we conduct a tiered analysis: one must have had COVID to develop long COVID, and by the structure of the survey, one must have had long COVID to experience activity limitations.

The regression analysis is based on a logistic function that relates the odds of having long COVID (or having activity limitations due to long COVID) to a set of individual characteristics and, in the full model, to a set of interaction terms. All regression results are presented as odds ratios because the interpretation of the effects is more intuitive than the logistic regression coefficients. Individual demographic characteristics are represented by dummy variables for being a woman, having at least a four-year college degree, and being a sexual or gender minority. Being a woman is proxied by being assigned female sex at birth rather than by self-identified gender because the survey response for the gender question is constructed as "female, male, transgender, and none of these." A woman/man binary variable inappropriately groups trans people, men, and genderqueer people together as men, and this construction effectively excludes trans and genderqueer people from the regression analysis. Hence we use female sex assigned at birth to retain transgender and genderqueer individuals in the regression and identify the variable as 'female' rather than woman. People who are SGM self-identified their sexual orientation as gay, bisexual, "something else," or their gender as transgender or genderqueer.[20] Our regression specification also comprises a set of mutually exclusive dummy variables for race/ethnicity (Hispanic, Black non-Hispanic, Asian non-Hispanic, and other race non-Hispanic), where white non-Hispanic is the reference group.

The complete model contains a full set of two-way and three-way interaction terms between being assigned female sex at birth, having at least a four-year degree, being a sexual and gender minority, and belonging to a historically underrepresented racial/ethnic group. Hence each intersectional identity (including two or three categories) has its own indicator variable, and altogether we have 18 two-way interaction terms and 16 three-way interaction terms. Our regressions also contain dummy variables specific to the survey month. These monthly fixed effects control for month-specific variation, monthly trends, and any unobserved heterogeneity. In particular, the fixed effects would capture other unobservable factors not captured in our repeated cross-section data that may influence having long COVID, change contemporaneously from month-to-month, and are common across individuals. For example, if the prevalence of long COVID trended downward over time due to newer strains of the virus being less likely to cause long-term symptoms, or due to protective effects from vaccination, this downward trend would be captured by the monthly fixed effects.

### C. Limitations

Our study uses data from a large sample weighted to be nationally representative and provide estimates of population prevalence, and there are both benefits and limitations. First, the HPS survey question about long COVID did not name long COVID specifically, which may help identify more people with long COVID. At least one study found considerable variation in symptom type, symptom severity, symptom attribution, and uncertainty about the applicability of the label "long COVID" among those experiencing persistent symptoms who were enrolled in a long COVID study, introducing potential bias because people do not perceive themselves as having "long COVID." [25] A potential limitation is that the retrospective variables we use may be subject to recall bias – the (in)ability of respondents to accurately report their significant life events retrospectively. However, because the survey was a rapid deployment survey that started in March 2020, we do not believe this type of bias would change our results in any meaningful way.

Another limitation is the small sample sizes of the sub-populations represented by some of the two-way and three-way interaction terms in the regression analysis (Appendix Table 1) The smallest groups are people who self-identified as Asian, Black, or other race who are also sex or gender minorities. When divided further by education, the groups with smallest number of respondents have 157, 168, and 265 people, respectively. The other small group, with 150 respondents, is people assigned female at birth who are sex or gender minorities and self-identified as Asian. A third potential limitation is that different populations have experienced different access to testing and different awareness of symptoms, and this impacts their ability to answer if they have ever had COVID. This issue could also affect the survey's measure of whether or not someone developed long COVID, particularly given public sentiment shifting toward denial of impact. To the extent that Hispanic and non-Hispanic Black individuals are more likely than Asian and non-Hispanic white individuals to suspect that they were infected [26], these sub-populations may be more likely to get tested and to report developing long COVID, if testing is accessible.

## III. Results

### A. Descriptive Analysis

The data indicate that an average of 31.1% of adults in the U.S. who had COVID (14.3% of all adults) developed long COVID. This is consistent with other studies in the U.S. and England. [23, 24] Prevalence and activity limitations in daily life from long COVID (Figure 1) are distributed along the lines of gender identity, education, ethnicity, and SGM status. Women, people without a four-year college degree, Hispanic people, and people who are SGM, especially those who identify as transgender, are more likely than their counterparts to experience long COVID (Panel A). People with a bachelor's degree and men have the lowest prevalence of long COVID (7.6% and 7.3%). The vast majority (80%) of people who had long COVID reported that they had activity limitations in daily life due to their symptoms, with particularly high rates for transgender individuals, non-white people, sexual and gender minorities, and individuals without a college degree (Panel B). Although women are considerably more likely than men to develop long COVID, they are slightly less likely than men to report activity limitations among those who reported having long COVID.

**Figure 1.** Prevalence of Long COVID and Activity Limitations Due to Long COVID, by Demographic Group

Panel A: Percentage of Adults Who Had COVID Who Developed Long Covid

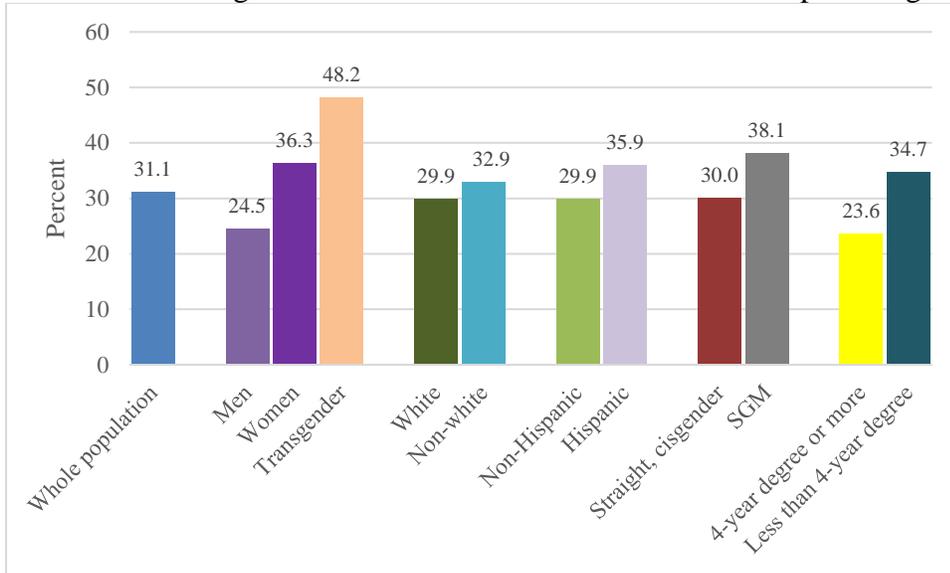

Panel B: Percentage of Adults with Long Covid Who Reported Activity Limitations

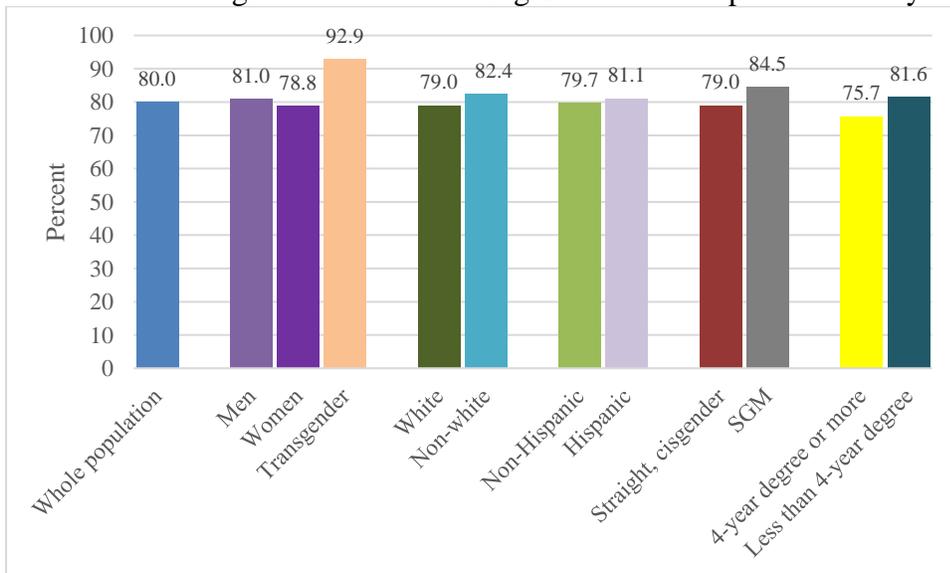

Our intersectional analysis in Figure 2 demonstrates that race, ethnicity, and education play major roles in the prevalence of long COVID and activity limitations due to long COVID among men and women, just as gender and education play important roles within racial and ethnic categories. Across all races, women were more prone than men to developing long COVID (Panel A). Panel A further shows a striking step-like pattern in which college-educated men had the lowest prevalence, followed by non-college educated men, then college-educated women, and lastly, with the highest prevalence, are women without BA/BS degrees. The figure also points to equalization in prevalence between white women with a college degree and men without a degree — and the lack of equalization in other race/ethnicity groups. When averaging together across race and ethnicity, overall, among all college-educated men, the prevalence of long COVID was 17.9% while it was 40.6% for women without a degree. Men without a BA/BS degree and women with a degree were virtually the same at 27.6% and 27.9%, a phenomenon driven by long COVID prevalence among whites.

**Figure 2.** Prevalence of Long COVID and Activity Limitations Due to Long COVID, Intersectional Analysis

Panel A: Percentage of Adults Who Had COVID Who Developed Long Covid

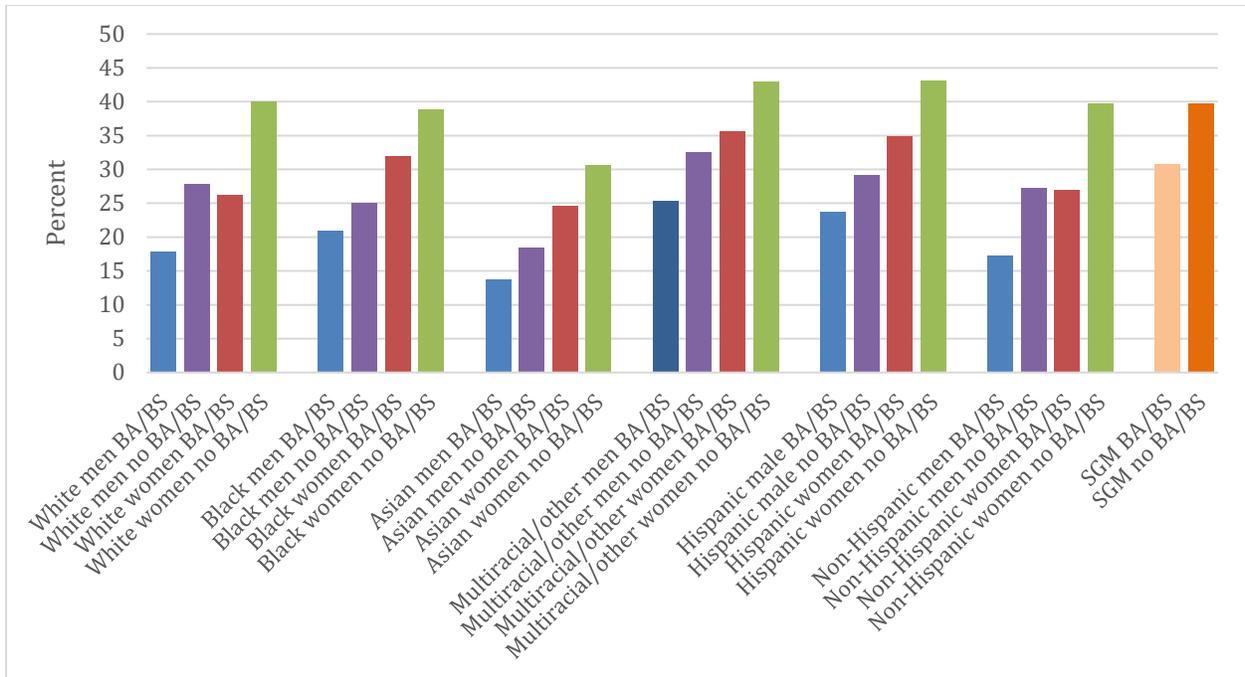

Panel B: Percentage of Adults with Long Covid Who Reported Activity Limitations

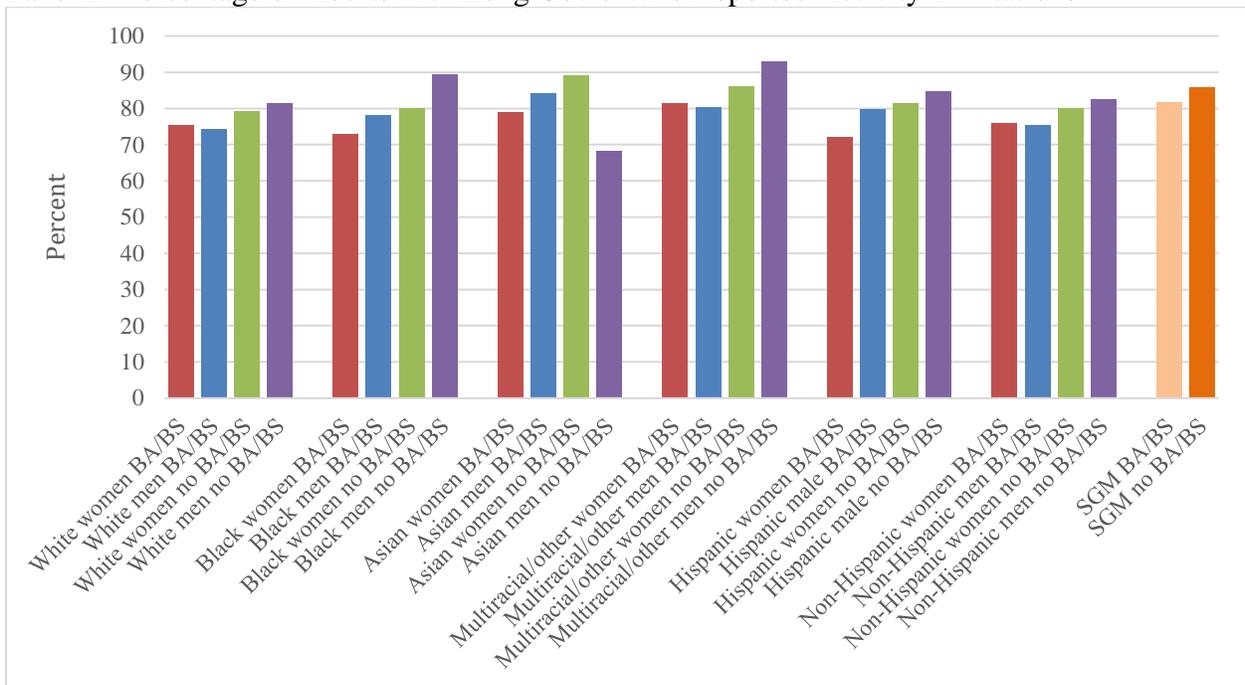

Source: Authors' analysis of data from the Household Pulse Survey, June 2022 to March 2023. Reported percentages are sample means.

Figure 2, Panel B shows that activity limitations do not follow the same pattern as prevalence. The relationships now appear to be more strongly delineated by education within racial and ethnic groups, while gender remains salient but the patterns for women and men are reversed. Women with college degrees tend to have lower rates of activity limitation, followed by men with college degrees, women without degrees, and finally men without degrees. The disparity is widest for women with college degrees and for men without college degrees among people who are Black, followed by Hispanic people, multiracial/other people, white people, and non-Hispanic people. The disparity in activity limitations by education is also apparent for sexual and gender minorities.

### B. Regression Analysis

Logistic regression results in Table 1, Column 1, indicate that major risk factors for long COVID are being assigned female sex at birth, not having a college degree, being a sexual or gender minority, being Hispanic, and being multiracial/other. In particular, among all adults who have had COVID, females have considerably greater odds (by a factor of 1.76) of having long COVID compared to males. Similarly, sexual and gender minority individuals have meaningfully greater odds (1.39) of getting long COVID compared to their heterosexual and cisgender counterparts. Among the race/ethnicity categories, the greatest risk of having long COVID is associated with being Hispanic and being multiracial/other, relative to being white non-Hispanic. In the opposite direction, higher education appears to have a protective effect. People who have at least a BA/BS degree have considerably lower odds (0.62) of having long COVID compared to people with less education. Also experiencing relatively lower odds of getting long COVID are Asian non-Hispanic individuals (0.71).

Column 2 of Table 1 reports results for the specification that includes the various permutations of two-way and three-way interaction terms. Our conclusions about being female, not having a college degree, and being a sexual or gender minority do not change. However, results for race and ethnicity are more nuanced. In particular, once we add in the interaction effects, the odds ratios for being Hispanic and for being multiracial/other are no longer statistically significant, while being Black non-Hispanic gains in statistical significance. There are a number of two-way interaction terms that have odds ratios that are both meaningful and statistically significant. We find particularly high odds of developing long COVID for Hispanic females, Black non-Hispanic females, Hispanic college graduates, Black college graduates, and multiracial/other college graduates. Although college may have a protective effect, that protective effect appears to only apply to white individuals. Also, among the two-way interactions, we see that Asian sexual and gender minorities have considerably lower odds of experiencing long COVID, although this result appears to hold only for those with less education as the three-way interaction term shows a very high odds ratio for Asian sexual and gender minorities with a college degree. None of the other three-way interaction terms have statistically significant odds ratios, except for multiracial sexual and gender minorities with a college degree, who have lower odds of developing long COVID.

Conditional on having long COVID, who is more likely to report activity limitations? Table 1, column 3 shows that females and sexual and gender minorities have greater odds of having activity limitations due to long COVID. However, the only racial/ethnic group to have greater odds of having activity limitations relative to white non-Hispanic people is the multiracial/other race group. Column 4 confirms that there are variations within the broader demographic categories. When we add the interaction effects, Hispanic females, college-educated females, and especially Asian sexual and gender minorities have greater odds of having activity limitations. All but one of the three-way interaction terms are statistically insignificant. Although Asian sexual and gender minorities with a college degree have a very high odds of experiencing long COVID, they have a particularly low odds of reporting activity limitations due to long COVID.

**Table 1.** Logistic Regression Results for Has Long Covid and Has Activity Limitations Due to Long COVID (Odds Ratios)

|  | Has Long Covid | | Has Activity Limitations | |
| --- | --- | --- | --- | --- |
|  | (1) | (2) | (3) | (4) |
| Has 4-yr degree | 0.617*** | 0.551*** | 0.959 | 0.858** |
|  | (0.009) | (0.016) | (0.029) | (0.054) |
| Female | 1.756*** | 1.701*** | 1.104*** | 0.991 |
|  | (0.032) | (0.047) | (0.042) | (0.056) |
| Sexual/gender minority | 1.390*** | 1.435*** | 1.442*** | 1.398** |
|  | (0.038) | (0.097) | (0.075) | (0.186) |
| Race/ethnicity (reference white non-Hispanic) | | | | |
| Hispanic | 1.163*** | 1.016 | 0.831*** | 0.741** |
|  | (0.032) | (0.062) | (0.047) | (0.097) |
| Black non-Hispanic | 0.973 | 0.789*** | 0.821*** | 0.741* |
|  | (0.029) | (0.064) | (0.049) | (0.127) |
| Asian non-Hispanic | 0.707*** | 0.685*** | 0.843* | 0.616* |
|  | (0.034) | (0.087) | (0.083) | (0.174) |
| Other race non-Hispanic | 1.230*** | 1.113 | 1.231*** | 1.157 |
|  | (0.048) | (0.097) | (0.087) | (0.196) |
| Age | 1.002*** | 1.001** | 1.016*** | 1.017*** |
|  | (0.001) | (0.001) | (0.001) | (0.001) |
| Two-way interaction terms | | | | |
| Female*Hispanic |  | 1.136* |  | 1.348* |
|  |  | (0.084) |  | (0.208) |
| Female*Black |  | 1.262** |  | 1.196 |
|  |  | (0.115) |  | (0.227) |
| Female*Asian |  | 1.052 |  | 1.339 |
|  |  | (0.184) |  | (0.478) |
| Female*Other race |  | 1.013 |  | 1.160 |
|  |  | (0.110) |  | (0.232) |
| Female*Has BA/BS |  | 0.986 |  | 1.158* |
|  |  | (0.036) |  | (0.088) |
| Female*SGM |  | 0.993 |  | 1.163 |
|  |  | (0.078) |  | (0.174) |
| Has BA/BS*Hispanic |  | 1.387*** |  | 1.099 |
|  |  | (0.118) |  | (0.199) |
| Has BA/BS*Black |  | 1.542*** |  | 0.929 |
|  |  | (0.175) |  | (0.229) |
| Has BA/BS*Asian |  | 1.068 |  | 1.379 |
|  |  | (0.152) |  | (0.444) |
| Has BA/BS*Other race | | 1.397*** | | 1.153 |
| | | (0.163) | | (0.262) |
| Has BA/BS*SGM | | 1.129 | | 1.112 |
| | | (0.093) | | (0.187) |
| SGM*Hispanic | | 1.059 | | 1.021 |
| | | (0.156) | | (0.325) |
| SGM*Black | | 0.718 | | 1.334 |
| | | (0.158) | | (0.685) |
| SGM*Asian | | 0.519** | | 3.547** |
| | | (0.165) | | (2.252) |
| SGM*Other race | | 1.265 | | 1.044 |
| | | (0.233) | | (0.350) |
| Three-way interaction terms | | | | |
| Female*Has BA/BS*SGM | | 0.980 | | 0.983 |
| | | (0.094) | | (0.189) |
| Female*Has BA/BS*Hispanic | | 0.973 | | 0.773 |
| | | (0.100) | | (0.166) |
| Female*Has BA/BS*Black | | 0.914 | | 1.183 |
| | | (0.116) | | (0.324) |
| Female*Has BA/BS*Asian | | 1.134 | | 0.858 |
| | | (0.218) | | (0.350) |
| Female*Has BA/BS*Other race | | 0.967 | | 0.805 |
| | | (0.139) | | (0.217) |
| SGM*Has BA/BS*Hispanic | | 0.976 | | 1.088 |
| | | (0.130) | | (0.282) |
| SGM*Has BA/BS*Black | | 1.150 | | 1.105 |
| | | (0.210) | | (0.430) |
| SGM*Has BA/BS*Asian | | 2.123*** | | 0.293** |
| | | (0.579) | | (0.165) |
| SGM*Has BA/BS*Other race | | 0.704** | | 0.948 |
| | | (0.123) | | (0.293) |
| Female*SGM*Hispanic | | 0.806 | | 0.588 |
| | | (0.130) | | (0.197) |
| Female*SGM*Black | | 0.924 | | 0.474 |
| | | (0.213) | | (0.243) |
| Female*SGM*Asian | | 1.004 | | 0.396 |
| | | (0.297) | | (0.239) |
| Female*SGM*Other race | | 0.807 | | 0.775 |
| | | (0.167) | | (0.295) |
| Constant | 0.274*** | 0.288*** | 0.213*** | 0.222*** |
| | (0.011) | (0.012) | (0.018) | (0.020) |
| Sample size | 285,749 | 285,749 | 59,166 | 59,166 |
| F Statistic | 164.01*** | 85.64*** | 23.91*** | 9.46*** |

Source: Authors' analysis of data from the Household Pulse Survey, June 2022 to March 2023.

Dummy for survey month in all regressions. Standard errors in parentheses. The notation [***] is p<0.01, [**] is p<0.05, * is p<.10. Sample for (1) and (2) consists of all adults who had COVID, and sample for (3) and (4) consists of all adults who had long COVID.

## IV. Discussion

### A. Main Findings

Our tiered analysis indicates that one in three people who had COVID developed long COVID, and four-in-five people who had long COVID had symptoms severe enough to impede their ability to carry out day-to-day activities. Long COVID is common, and once a person has long COVID, the vast majority have debilitating symptoms. These overall conclusions are depicted in Figure 3.

**Figure 3.** Tiered Representation of Overall Findings

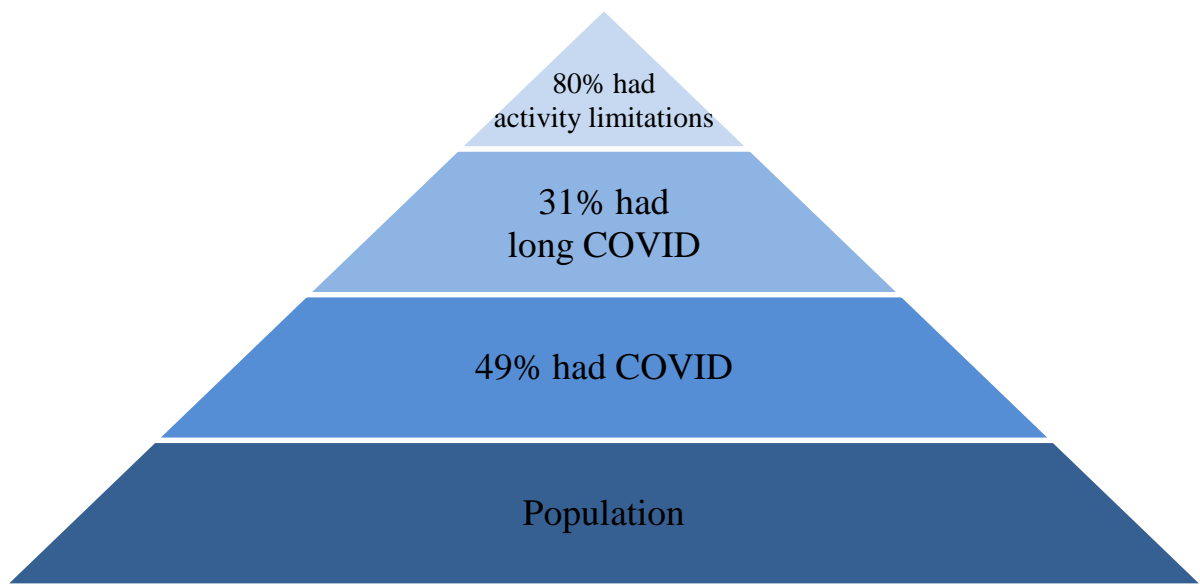

Our results further show that, in aggregate, long COVID prevalence conceals substantial variation across demographic groups: prevalence is statistically significantly higher among females, people without four-year college degrees, and sexual and gender minorities. These results also hold in the full model for the odds of developing long COVID that includes detailed interaction terms. Yet when we consider activity limitations among people with long COVID, the variations among demographic groups are less pronounced: most people who had long covid – 80% – had activity limitations. The disparity between women and men almost disappears and the protective effect of a college education is greatly diminished (although it is statistically significant in the model with the interaction terms). In other words, demographics affect the prevalence of long COVID, but once a person has long COVID, it is highly likely that their symptoms are debilitating enough to reduce the quality of life by curtailing the ability to carry out day-to-day activities.

Our other key finding is that the role of race/ethnicity in explaining the prevalence of long COVID is nuanced because of intersectional effects. In our descriptive analysis and the simple regression model, individuals who are Hispanic or multiracial/other are more likely to develop long COVID, but once we add two-way and three-way interaction effects, we see that these broad effects for Hispanic and multiracial/other individuals are no longer statistically significant. Rather, intersections of race/ethnicity with gender, education, and SGM status matter more. For example, Black and Hispanic women have significantly higher odds of developing long COVID compared to all non-Black and non-Hispanic individuals and compared to Black and Hispanic men. Moreover, having a college degree is only protective for people who identify as white or as straight/cisgender. Asian sexual and gender minorities with college degrees have particularly high odds of developing long COVID, but it is not clear why.

On their own and intersectionally, gender, education, being a sexual or gender minority, and race/ethnicity are all salient in predicting who develops long COVID, and to some extent – but less so – in predicting who develops activity limitations due to long COVID symptoms. These overall results are in line with other studies using different samples to examine social determinants of health as predictors of long COVID. In a major study using data from England,

Whitaker et al. (2022) find that being older, a woman, overweight or obese, low-income, a healthcare or home health aide, in the bottom two quintiles of a deprivation index, a current smoker, having a prior hospitalization with COVID, and being a current vaper were, in that order, are the strongest predictors of persistent symptoms. [23] The authors conclude that indicators associated with lower socioeconomic status are particularly important in explaining long COVID prevalence rates over and above the disparities across demographic groups already observed in morbidity and mortality from COVID-19.

**B. Inequalities, Inequities, and Determinants of Health**

Inequalities in COVID outcomes are related to disparities in other chronic diseases and have a complex relationship with biological and social determinants of health. [27] In the case of gender, women on average have higher morbidity rates than men, while men have shorter life expectancies and higher mortality rates than women, a disparity explained not only by biological factors (such as differences in immune systems and in hormones) but also social factors (such as gender differences in healthcare seeking behaviors). [28] Health inequities by socioeconomic status can also be explained by both biological factors (such as chronic stress of economic and psychological deprivation leading to immunosuppression) and social factors (such as insufficient access to quality healthcare) and the combination thereof. [28, 29]

Other biosocial factors relevant to explaining long COVID prevalence are pre-existing chronic health conditions, as people with chronic conditions such as diabetes, hypertension, and obesity are at increased risk of severe illness and complications from COVID-19. Physiological factors play a role in explaining long COVID, although evidence indicates that these comorbidities, too, are subject to social determinants of health. These conditions are often more prevalent among marginalized populations due to historical context and related, persistent inequalities/inequities that limit access to healthy food, safe living environments, and quality healthcare.

Disparities by gender, socioeconomic status, and race/ethnicity are widely observed in health outcomes, including COVID-19 morbidity and mortality, and are often explained by social determinants of health. [30] Closely connected to chronic health conditions are access to health care and health insurance. Individuals with economic and social vulnerabilities, especially those with chronic conditions, are typically at greater risk of exposure; they confront barriers to timely and effective healthcare; they have limited access to testing, treatment, and preventive measures; and they face challenges in following public health recommendations. For example, it is difficult for a parent with caregiving responsibilities to quarantine in a small home; and for the unhoused, washing one's hands can be a major challenge. These social problems put a disproportionate burden of COVID-19 on already-vulnerable populations.

This argument also applies to people who are socially disadvantaged, especially transgender individuals, almost half of whom developed long COVID from a COVID infection and over 90% had activity limitations, as shown in our analysis. Their heightened risk largely reflects the well-known disparities that sexual and gender minority individuals face in accessing healthcare. [31, 32]

### C. Embodiment at Work

Embodiment is an ontogenetic process that includes generational changes through epigenetics, in which past environmental factors may change biology and current environmental factors. [33] Biological constraints and possibilities combine with the ways in which capital organizes production and reproduction to simultaneously structure pathways to embodiment. [20, 34 ]. The capitalist organization of production and reproduction allows – even compels – rapid circulation of people, products, and pathogens across long distances. Ecosocial theory's embrace of the political economy of health and its materialist foundations suggest that labor processes merit particular attention. Work is not separable from some chronic conditions; indeed the centrality of work and consequent economic security contributes to many biological manifestations of social relations. Historically, chattel slavery, forced labor, indentured servitude, and child labor are implicated in immunodeficiency. [35] Likewise, the spread of pathogens is concomitant with the movement of people, evident in the spread of smallpox along slave trade networks in which indigenous people were forced into colonial production and reproduction. [34, 36]

In a viral challenge study, Cohen et al. (1998) found that chronic stressors related to work, especially unemployment and underemployment, were the strongest risk factors for developing a rhinovirus-induced cold. [37] Further, exposure to stressful events is more common in low socioeconomic status neighborhoods. [38] Some scholars expected gender roles to mean that men experienced more stress related to paid work and women experienced more interpersonal stress related to care work. [38] Evidence indicates that stressful events *are* gendered, but not as anticipated: a meta-analysis of 119 studies between 1960 and 1996 found that women report greater exposure to stressful events than men in both work and interpersonal domains [38].

As in previous pandemics, observed disparities in long COVID appears to have a great deal to do with gendered and racialized divisions of labor. Work is so fundamental to embodiment that disparities in long COVID seem to revolve around occupational segregation in the labor market and the household division of labor. Certain populations, especially women and those with lower socioeconomic status, are disproportionately represented in essential occupations that require close contact with others, increasing the risk of repeated exposure to the virus and the need to continue working even when sick. [39]

Gender plays a critical role in economic security in part because it helps determine the kinds of work available to women and men. In particular, women are overrepresented in the care sector and women of color are crowded into poorly paid care occupations, like personal care aides and home health aides. [40] Many women work in direct care occupations that mimic – in the form of tasks and their perceived value – unpaid work in the household division of labor. The division of household labor is structured by the same gender roles that influence which demographic groups do which kinds of paid work. Occupational segregation and the gender division of labor in the household increase economic insecurity and distress among women; these are well-documented phenomena. [41] Not coincidentally, they also heighten the risk of COVID infection and reinfection because of the intimate nature of work in both sites.

Risks related to occupational exposure and gender roles may account for some disparity in prevalence through reinfection. Although scientific evidence about the relationship between reinfection and the risk of developing long COVID is mixed, reinfection generates greater risk for long COVID than successfully avoiding reinfection. [42, 43] In principle, having a white-collar job that one can do from home should help to avoid reinfection, but remote work may not be a protective factor for women if they are at risk of household exposure through their care responsibilities. A 2023 study found that 70 percent of within-household transmission of COVID originated with schoolchildren, meaning that women who care for ill children and other family members are at high risk of exposure at home. [44]

Also intertwined in these predictors of inequities in long COVID outcomes are structural racism and discrimination. Marginalized populations, especially racial/ethnic minorities and sexual and gender minorities, often face systemic barriers that affect their health outcomes. These factors are evident in both chronic diseases and the disparate impacts of COVID-19. Race and ethnicity have historically contingent constructions that have been used to channel women of color into certain occupations. [45] Enduring structural, institutional, and medical racism, along with interpersonal discrimination, all point to ways that labor – paid and unpaid – is implicated in creating and reinforcing racialized inequities and economic inequality. Where the prevalence of long COVID was higher among people of color, it may be a function of inequity in access to care, paid sick leave, and remote work. Like other studies that disaggregate by race [4], we find fairly small Black/white disparities, possibly due to confounding factors in the data. Survivorship bias is one such factor. People who might have had symptoms of long COVID if they had survived, but did not, leads to a lower estimated prevalence of long COVID among Black and other marginalized populations, the populations worst impacted by COVID. [46]

In closing, long COVID is an urgent issue presenting ongoing challenges to already-disadvantaged people, with implications for individuals, families, and communities. It can have lasting physical, mental, emotional, and neurological effects that tend to impair one's ability to engage in day-to-day activities. Further, because occupational segregation concentrates women of color and women without college degrees in care industries, long COVID has indirect impacts on recipients of care. Brain fog among people who distribute medications, for example, could have high human costs. History suggests that occupational segregation and, to a limited degree, the household division of labor can be addressed through interventions. Policies that help improve employment opportunities in a wider range of occupations for women are critical. Employers could provide benefits and programs more amenable to the realities of the household division of labor, including on-site childcare, remote work options, flexible scheduling, and paid family leave. Where profit- and revenue-maximizing employers resist implementing such policies, the state can provide resources itself. More broadly, delinking survival and the labor market through a stronger social safety net could help raise the perceived value of care and caregivers as well. An economy in which many of the least secure essential workers are those most burdened by long COVID may prove as unsustainable as it is undesirable.

**Declarations**

Ethics approval and consent to participate: Not applicable

Consent for publication: Not applicable

Availability of data and materials: The datasets generated and/or analyzed during this study are publicly available through https://www.census.gov/data/experimental-data-products/household-pulse-survey.html. The aggregated dataset is available in the Figshare repository, [weblink to be added]

Competing interests: The authors declare that they have no competing interests

Funding: Yana van der Meulen Rodgers was supported in part by a grant from the National Institute on Disability, Independent Living, and Rehabilitation Research (NIDILRR) for the Rehabilitation Research & Training Center (RRTC) on Employment Policy: Center for Disability-Inclusive Employment Policy Research Grant [grant number #90RTEM0006-01–00] and by the RRTC on Employer Practices Leading to Successful Employment Outcomes Among People with Disabilities, Douglas Kruse PI, Grant [grant number #90RTEM0008-01-00].

Acknowledgements: The authors thank Doug Kruse, Lisa Schur, Mason Ameri, Samantha Deane, Michael Morris, Nanette Goodman, and two anonymous reviewers for their constructive feedback.

Authors' information:
JC is an Associate Professor at Miami University, Ohio, USA and a joint researcher with Ezintsha, Faculty of Medicine, University of the Witwatersrand. Her mixed-methods research focuses on work, race and gender inequities, political economy of health, and social determinants of health. She earned her PhD in Economics at the University of Massachusetts in 2012.

YR is a Professor in the Department of Labor Studies and Employment Relations, and in the Department of Women's and Gender Studies, at Rutgers University. She also serves as Faculty Director of the Center for Women and Work at Rutgers. Yana specializes in using quantitative methods and large data sets to conduct research on women's health, labor market status, and well-being.

Appendix Table 1.

Sample Means and Cell Sizes

| | Adults who had Covid | | Adults who had long Covid | |
|---|---|---|---|---|
| | Proportion | Number of respondents | Proportion | Number of respondents |
| Has long Covid | 0.28 | 79,401 | 1.00 | 59,166 |
| Has activity limitations | 0.10 | 27,678 | 0.36 | 21,427 |
| Has 4-yr degree | 0.56 | 160,745 | 0.46 | 27,229 |
| Female | 0.59 | 167,784 | 0.68 | 40,447 |
| Sexual/gender minority | 0.12 | 33,714 | 0.15 | 8,875 |
| Race/ethnicity | | | | |
| White | 0.75 | 215,499 | 0.74 | 43,567 |
| Hispanic | 0.10 | 27,826 | 0.11 | 6,700 |
| Black non-Hispanic | 0.06 | 18,239 | 0.07 | 4,075 |
| Asian non-Hispanic | 0.04 | 12,285 | 0.03 | 1,663 |
| Other race non-Hispanic | 0.04 | 11,900 | 0.05 | 3,161 |
| Age | 49.32 | 285,749 | 48.82 | 59,166 |
| Two-way interaction terms | | | | |
| Female*White | 0.44 | 124,329 | 0.49 | 29,191 |
| Female*Hispanic | 0.06 | 17,130 | 0.08 | 4,733 |
| Female*Black | 0.05 | 13,054 | 0.06 | 3,318 |
| Female*Asian | 0.02 | 5,905 | 0.02 | 981 |
| Female*Other race | 0.03 | 7,366 | 0.04 | 2,224 |
| Female*Has BA/BS | 0.32 | 90,172 | 0.30 | 17,955 |
| Female*SGM | 0.07 | 21,116 | 0.11 | 6,319 |
| Has BA/BS*White | 0.44 | 125,759 | 0.35 | 20,634 |
| Has BA/BS*Hispanic | 0.04 | 11,738 | 0.04 | 2,514 |
| Has BA/BS*Black | 0.03 | 8,156 | 0.03 | 1,646 |
| Has BA/BS*Asian | 0.03 | 9,609 | 0.02 | 1,180 |
| Has BA/BS*Other race | 0.02 | 5,483 | 0.02 | 1,255 |
| Has BA/BS*SGM | 0.07 | 18,837 | 0.07 | 4,242 |
| SGM*White | 0.08 | 24,137 | 0.11 | 6,298 |
| SGM*Hispanic | 0.02 | 4,330 | 0.02 | 1,239 |
| SGM*Black | 0.01 | 1,836 | 0.01 | 414 |
| SGM*Asian | 0.00 | 1,256 | 0.00 | 231 |
| SGM*Other race | 0.01 | 2,155 | 0.01 | 693 |
| Three-way interaction terms | | | | |
| Female*Has BA/BS*SGM | 0.04 | 11,288 | 0.05 | 2,930 |
| Female*Has BA/BS*White | 0.25 | 70,066 | 0.23 | 13,415 |
| Female*Has BA/BS*Hispanic | 0.02 | 6,794 | 0.03 | 1,712 |
| Female*Has BA/BS*Black | 0.02 | 5,578 | 0.02 | 1,299 |
| Female*Has BA/BS*Asian | 0.02 | 4,472 | 0.01 | 676 |

| | | | |
|---|---|---|---|
| Female*Has BA/BS*Other race | 0.01 | 3,262 | 0.01 | 853 |
| SGM*Has BA/BS*White | 0.05 | 14,229 | 0.05 | 3,144 |
| SGM*Has BA/BS*Hispanic | 0.01 | 1,927 | 0.01 | 508 |
| SGM*Has BA/BS*Black | 0.00 | 821 | 0.00 | 168 |
| SGM*Has BA/BS*Asian | 0.00 | 910 | 0.00 | 157 |
| SGM*Has BA/BS*Other race | 0.00 | 950 | 0.00 | 265 |
| Female*SGM*White | 0.05 | 15,074 | 0.08 | 4,465 |
| Female*SGM*Hispanic | 0.01 | 2,673 | 0.01 | 860 |
| Female*SGM*Black | 0.00 | 1,271 | 0.01 | 332 |
| Female*SGM*Asian | 0.00 | 647 | 0.00 | 150 |
| Female*SGM*Other race | 0.01 | 1,451 | 0.01 | 512 |
| Sample Size | 285,749 | 285,749 | 59,166 | 59,166 |